\documentclass[aip,preprint]{revtex4-1}

\draft 
\usepackage{graphicx}
\usepackage{bm}
\usepackage{amsfonts}
\usepackage{amsmath}
\usepackage{natbib}
\usepackage{epstopdf}
\draft 

\begin{document}


\title{The hydraulic bump: the surface signature of a plunging jet} 


\author{M. Labousse}
\affiliation{Institut Langevin, ESPCI ParisTech,\\ 1 rue Jussieu 75005 Paris, France, EU}
\author{J.W.M Bush}%
\email[]{bush@math.mit.edu}
\affiliation{ 
Department of Mathematics, Massachusetts Institute of Technology,\\ 77 Massachusetts Avenue, Cambridge, MA 02139, USA
}%

\date{\today}

\begin{abstract}
When a falling jet of fluid strikes a horizontal fluid layer, a hydraulic jump arises downstream of the point of impact provided a critical flow rate is exceeded. 
We here examine a phenomenon that arises below this jump threshold, a circular deflection of relatively small amplitude
on the free surface, that we call the hydraulic bump. The form of the circular bump can be simply understood in terms of the underlying 
vortex structure and its height simply deduced with Bernoulli arguments. As the incoming flux increases, a breaking of 
axial symmetry leads to polygonal hydraulic bumps. The relation between this polygonal instability and that 
arising in the hydraulic jump is discussed. The coexistence of hydraulic jumps and bumps can give rise to striking 
nested structures with polygonal jumps bound within polygonal bumps. The absence of a pronounced surface signature 
on the hydraulic bump indicates the dominant influence of the subsurface vorticity on its instability.

\end{abstract}

\maketitle

\section{\label{Introduction}Introduction}
\begin{figure*}
    \centering
    \includegraphics[width=160mm,height=120mm]{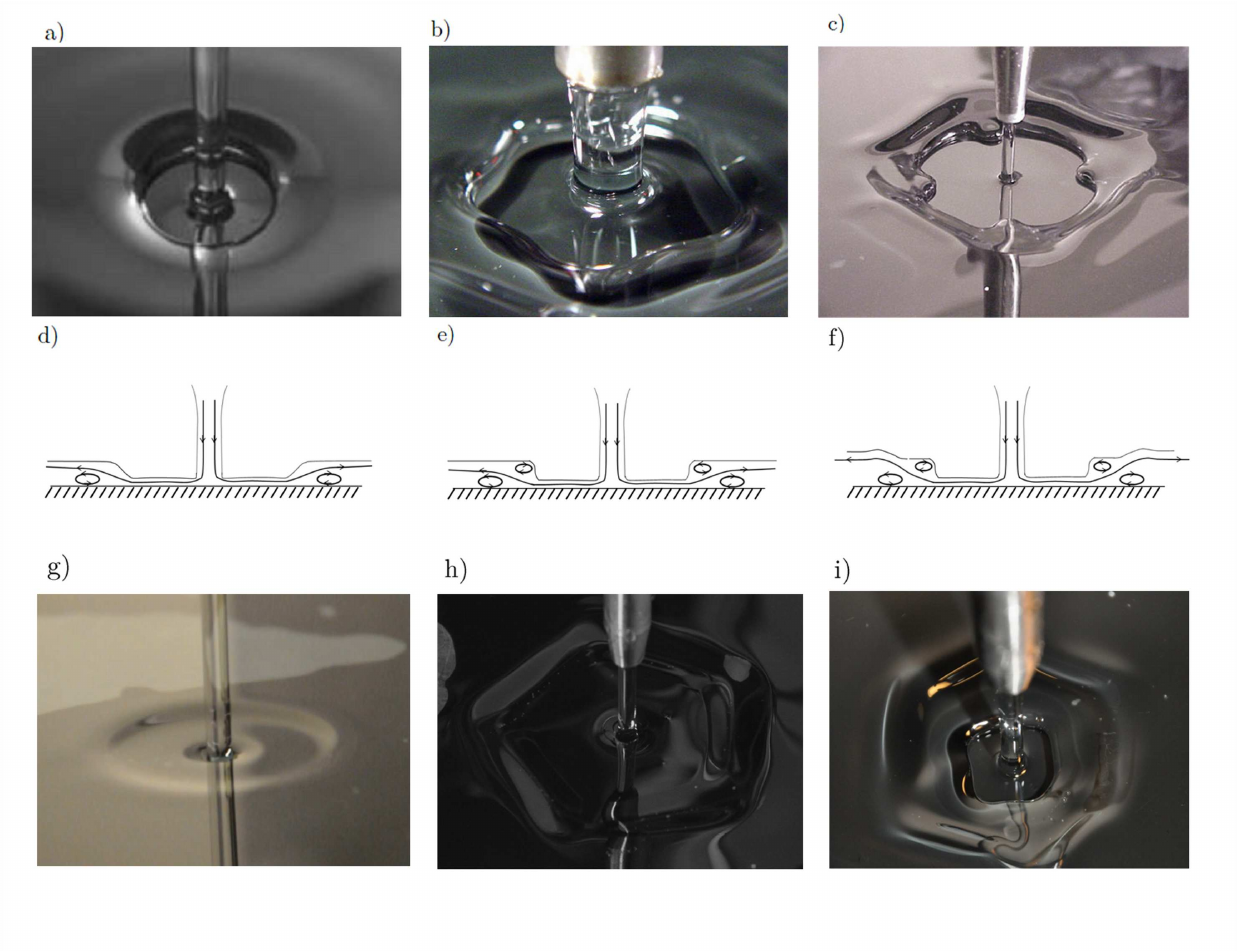}
\caption{a) The circular hydraulic jump \cite{Bush1}. b) A pentagonal hydraulic jump \cite{Bush2}. c)
A clover-shaped jump inside a square bump \cite{Bush2}. d)-f) Schematics illustration  of the hydraulic jump of type I d), IIa e) and IIb f). 
g) A circular hydraulic bump. h) A pentagonal hydraulic bump. i) A square 
hydraulic jump inside a hexagonal bump. a) Bush, J.W.M. and  Aristoff, J.M., J. Fluid Mech. \textbf{489}, (2003) reproduced with permission. b) and c) Bush, J.W.M. and Aristoff, J.M. and Hosoi, A.E., J. Fluid Mech. \textbf{558}, (2006) reproduced with permission}
\label{bumppicture}
\end{figure*}

When a falling jet of fluid strikes a horizontal fluid layer, several flow regimes may arise. The most distinctive 
phenomenon, the hydraulic jump, arises above a critical flow rate, and consists of a large-amplitude increase 
in fluid depth at a critical distance from the site of jet impact (Figure \ref{bumppicture} a). The circular hydraulic jump was first reported by 
B\'{e}langer \cite{Belanger} and Rayleigh \cite{Rayleigh14}, and subsequently studied  theoretically and experimentally by a number of investigators (see \cite{Tani48,Watson,Bohr93,Bohr96,Bohr97,Bush1,Kasimov08} and references therein).\\

Bohr \textit{et al.} \cite{Bohr96} and Watanabe \textit{et al.} \cite{Bohr03} distinguished between circular hydraulic jumps of type I and II. The type I jump (see Figures \ref{bumppicture} a, d) exhibits a single toroidal vortex downstream of the jump, henceforth ``primary vortex". As the outer depth is increased progressively, a separation of this vortex \cite{Bohr10} is observed, giving rise to a surface roller, henceforth ``secondary vortex" and a type II jump (Figure \ref{bumppicture}e,f). Yokoi \textit{et al.} \cite{Xiao02} presented a numerical investigation of the link between this vortex dynamics and the  underlying pressure distribution in the type II jumps and remarked upon the importance of surface tension in the transition from type I to II. The type II jumps are further classified \cite{Lienhard93} according to whether there is a substantial change in surface elevation downstream of the jump: if not, the jump is referred to as type IIa (Figure \ref{bumppicture} b and e); if so, type IIb (Figure \ref{bumppicture} c and f). Andersen \textit{et al.} \cite{Bohr10} and Bush \textit{et al.} \cite{Bush2} also reported the emergence of double jump structures in certain parameters regimes, wherein the free surface is marked by two discrete changes in depth.\\   

Remarkably, in certain parameter regimes, the circular hydraulic jump becomes unstable to polygons (Figure \ref{bumppicture}b), a phenomenon first reported by Ellegaard \cite{Bohr98,Bohr99}, and subsequently examined by Bohr and coworkers \cite{Bohr10,Bohr12} and Bush \textit{et al.} \cite{Bush2}.  Watanabe \textit{et al.} \cite{Bohr03} noted that the polygonal jumps arise exclusively with type II jumps, that is, when both primary and secondary vortices are present. Bush \textit{et al.} \cite{Bush2} highlighted the importance of surface tension in the polygonal instability of such jumps, suggesting that a modified Rayleigh-Plateau-like instability might be responsible. By considering a balance between the viscous stresses associated with the secondary vortex and the hydrostatic and curvature pressure, Martens \textit{et al.} \cite{Bohr12} developed a theoretical model for the jump shape that yields polygons similar to those observed experimentally. When surface tension dominates, they demonstrate that the wavelength of the instability is consistent with that of Rayleigh-Plateau. Nevertheless, they did not consider the potentially destabilizing influence of the pressure induced by the secondary roller vortex.  \\ 

Plateau\cite{Plateau73} examined the capillary pinch-off  of a fluid jet into droplets, a theoretical description of which was provided by Rayleigh \cite{Rayleigh79}. This Rayleigh-Plateau instability was extended to the case of a rotating fluid jet by several investigators,\cite{Hocking59,Pedley1,Weidman06} who demonstrated that the destabilizing influence of surface tension is enhanced by fluid inertia. While vortex rings were initially thought to be indestructible \cite{Kelvin67,Thomson83}, subsequent experimental, theoretical \cite{Widnall77,Maxworthy77,Saffman78} and numerical \cite{Ghoniem90} studies indicate that they are unstable to azimuthal wavelength disturbances at high Reynolds numbers, resulting in polygonal forms. We here explore the possible relevance of a such instabilities to the stability of the hydraulic jump and bump. \\
 
Bush \textit{et al.} \cite{Bush2} briefly mentioned the emergence of polygonal forms in the absence of hydraulic jumps, when a jet plunges into a relatively deep fluid. Perrard \textit{et al.} \cite{Perrard} recently reported that a heated toroidal fluid puddle bound in a circular channel and levitated via the Leidenfrost effect is also susceptible to polygonal instabilities. The axial symmetry breaking only arises in the presence of poloidal convection within the torus, again suggesting the importance of the vortical motion on the mechanism of instability.  \\

We here report a phenomenon that occurs well below the hydraulic jump threshold, when the free surface is only weakly perturbed by the plunging jet. When the fluid layer is sufficiently deep, 
a small-amplitude circular deflection arises at the free surface, a phenomenon that we christen the
hydraulic bump (Figure~\ref{bumppicture} g). As is the case for the hydraulic jump, as the incoming 
flux increases, the bump radius expands until a breaking of axial symmetry results in polygonal 
forms (Figure~\ref{bumppicture} h). In \S II, we report the results of our experimental investigation, and describe the flows observed. We rationalize the radius of the bump via simple scaling laws. Finally, in \S III we explore the 
connection between the polygonal instabilities on the hydraulic bump and their counterparts on the hydraulic jump. 

\section{\label{Experiments}Experiments}
\begin{figure*}
    \centering
\includegraphics[width=160mm,height=65mm]{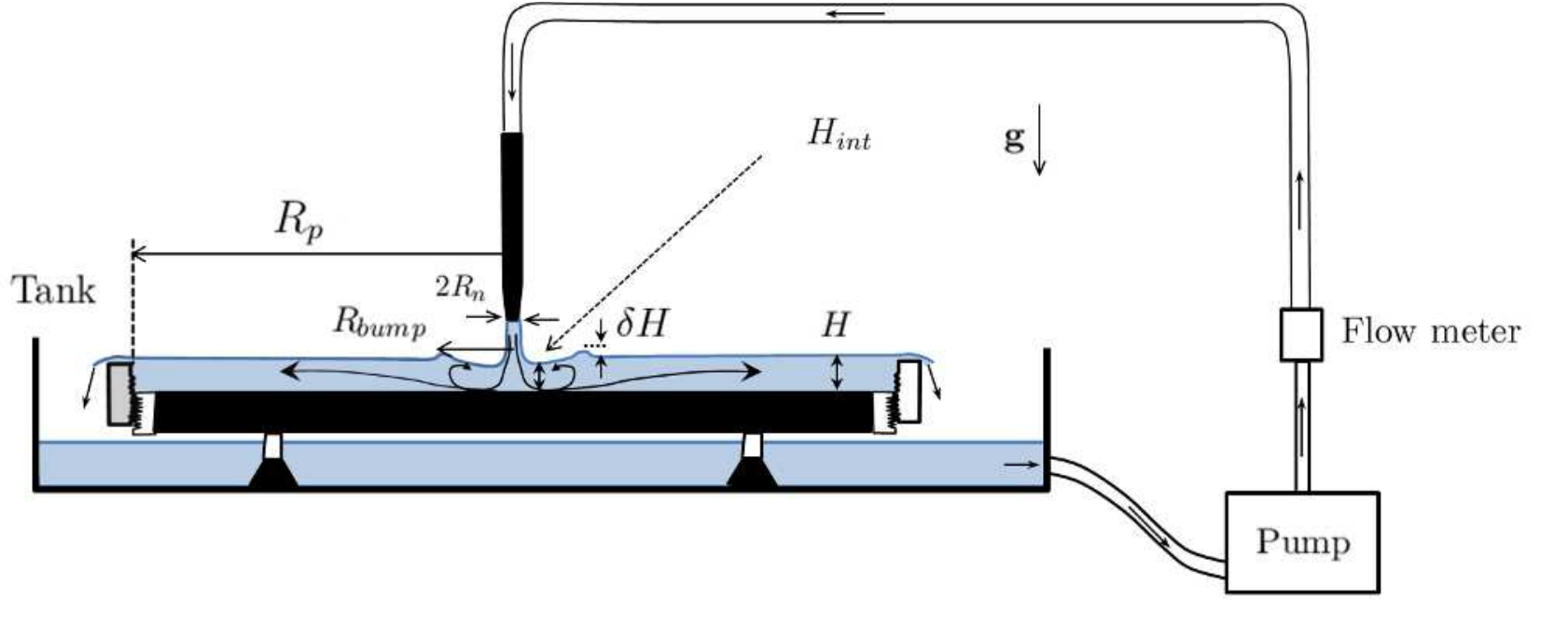}
\caption{Schematic illustration of the experimental apparatus and the hydraulic bump}
\label{experiment}
\end{figure*}

The experimental apparatus is shown in figure~\ref{experiment}. A glycerine-water solution with density $\rho$, 
kinematic viscosity $\nu$, and a surface tension $\gamma$ is pumped from the tank through a flow meter and a 
source nozzle of radius $R_n = 2.5$ mm. The resulting jet has a flux $Q$ and a radius at impact 
$r_j$ that differs from $R_n$, and varies weakly with flow rate and height in a manner detailed by Bush and Aristoff \cite{Bush1}. The jet
impacts the center of a flat plate of radius $R_p = 16.8$ cm surrounded by an outer wall whose height can be 
adjusted in order to control the outer fluid depth $H$. Radial gradations on the base plate indicate $0.5$ cm increments. 
Special care is taken to level the plate by adjusting its three supports and measuring the level along two perpendicular 
directions. The plate is horizontal to within $\pm \: 0.1^{\circ}$. We note that the flow structure is extremely sensitive to the levelling of the plate; indeed, an inclination of $1-2\:^{\circ}$ completely destroys the polygonal bump and jump forms.\\ 

The working fluid is a glycerine-water solution with viscosity ranging from $58$ to $96$ cS.
During the course of the experiments, water was added to compensate for evaporative losses. 
For the fluids considered, surface tension is roughly constant and equal to $68$ mN.m$^{-1}$. The average depth is determined by measuring 
the volume $V_t$ above the impact plate, which is known with a precision $\pm\: \delta V_t =2$ ml. 
Typically, $V_t \simeq 500$ ml and $H \simeq 5$ mm , so the error in depth 
$( \delta V_t H ) /V_t \simeq 0.01$ mm,  is sufficient for our experiments 
and smaller than would arise from a direct measurement.    
We visualize the flow structure by injecting submillimetric bubbles into the jet inlet with a syringe and taking photos that yield streak images of the bubble circulation. In passing through the pump and the flowmeter, these bubbles are generally
fractured into microbubbles that do not appreciably perturb the flow.  We denote by $R_{bump}$ the bump radius, $\delta H$ its height and $H_{int}$ the height just upstream of the bump. We denote the fluid velocity by $\mathbf{v}$ and its speed by $v$. This suggests the introduction of the Reynolds number $Re=v_j H/\nu$, with the jet speed $v_j=Q/(\pi R_n^2)$ being evaluated at the nozzle output, and a local Weber number $We=\rho Q^2 /( \gamma \pi^2 H R^2)$, with $R$ being the radius of the jump or the bump. \\
 
\begin{figure*}
    \centering

\includegraphics[width=160mm,height=60mm]{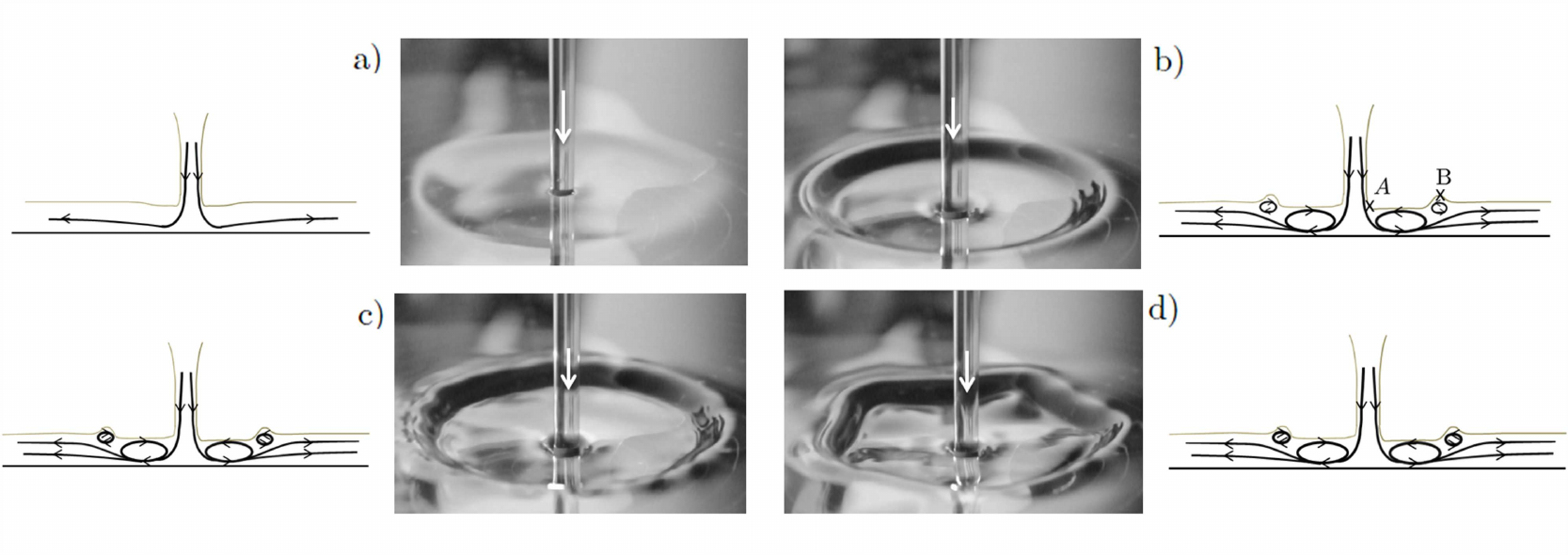}

\caption{Evolution of the flow generated by a plunging jet with increasing $Q$. Here, $\nu=68$ cS, $H=6.65$ mm.
Accompanying schematic illustrations of the subsurface flows induced from streak images are displayed. 
a) Below the bump transition ($Q=6.6$ ml.s$^{-1}$). b) The circular bump is marked by a toroidal vortex  
($Q=13.3$ ml.s$^{-1}$). c-d) At a critical flux ($Q=18.3$ ml.s$^{-1}$), the bump becomes susceptible to 
instabilities that result in a polygonal form.}
\label{medium}
\end{figure*}
Figure \ref{medium} illustrates the evolution of the flow generated by a plunging jet as the flux increases and the fluid depth $H$ is held constant. We note that the flux of the impacting jet is not sufficiently high to entrain air \cite{Bin93,Quere04}. Initially (Figure~\ref{medium}a), the plunging jet induces a slight circular deflection, 
perceptible only from an oblique angle, and the subsurface flow is predominantly radial. At a critical
flow rate, a recirculation eddy emerges, and with it the hydraulic bump (figure \ref{medium}b). We note that this subsurface recirculation eddy, or primary vortex, is accompanied by a small corotating secondary vortex with a surface signature that corresponds to the bump. As the flux increases, the bump increases in both amplitude and radius. At a critical
flux, azimuthal instabilities develop along its perimeter (figure \ref{medium}c), giving rise to a stable polygonal 
bump (figure \ref{medium}d). As the bump has a very modest surface 
signature, much less than the jump, we infer that the subsurface vortical structure is critical in its instability.\\

\begin{figure*}
    \centering
\includegraphics[width=95mm,height=70mm]{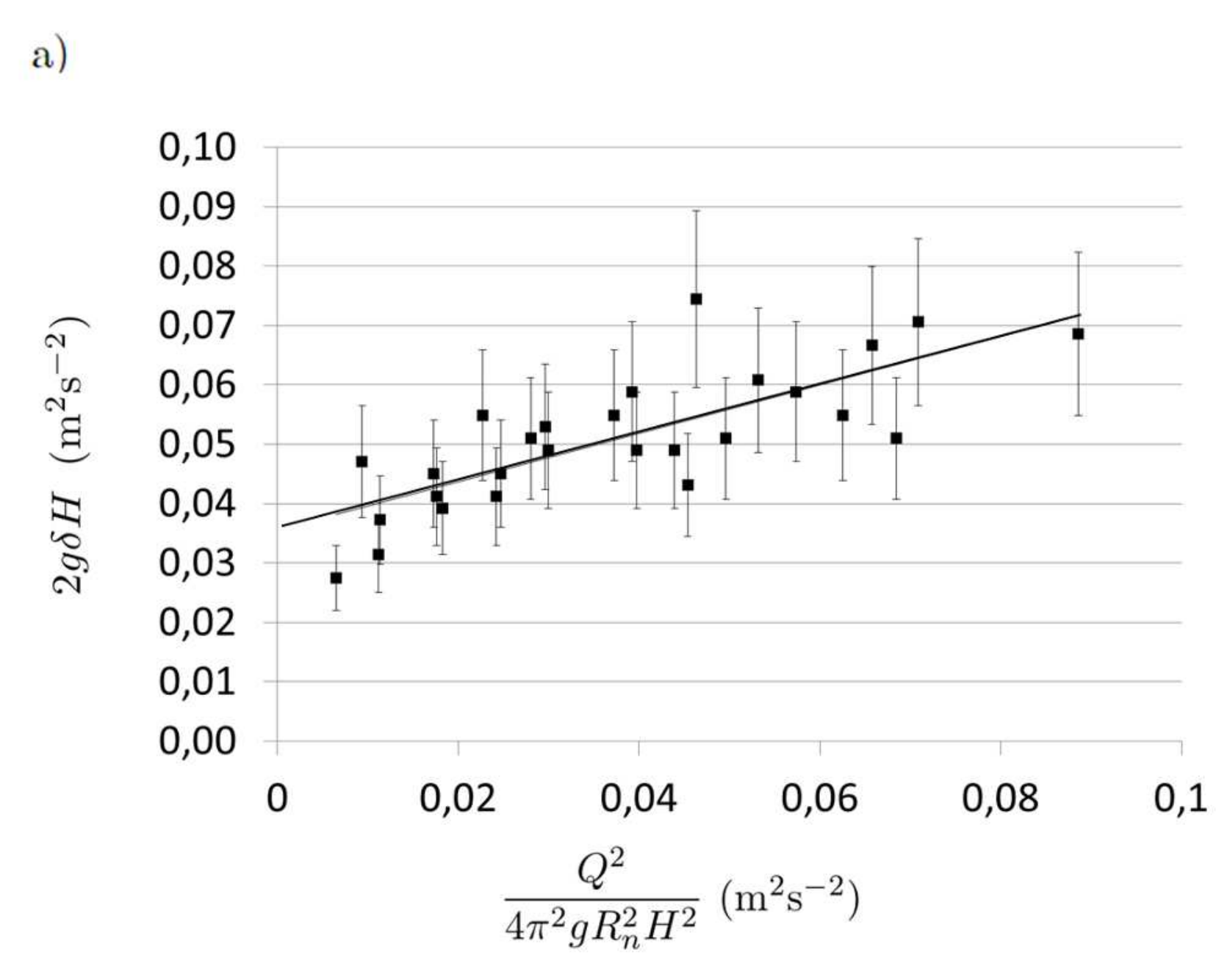}
\includegraphics[width=95mm,height=70mm]{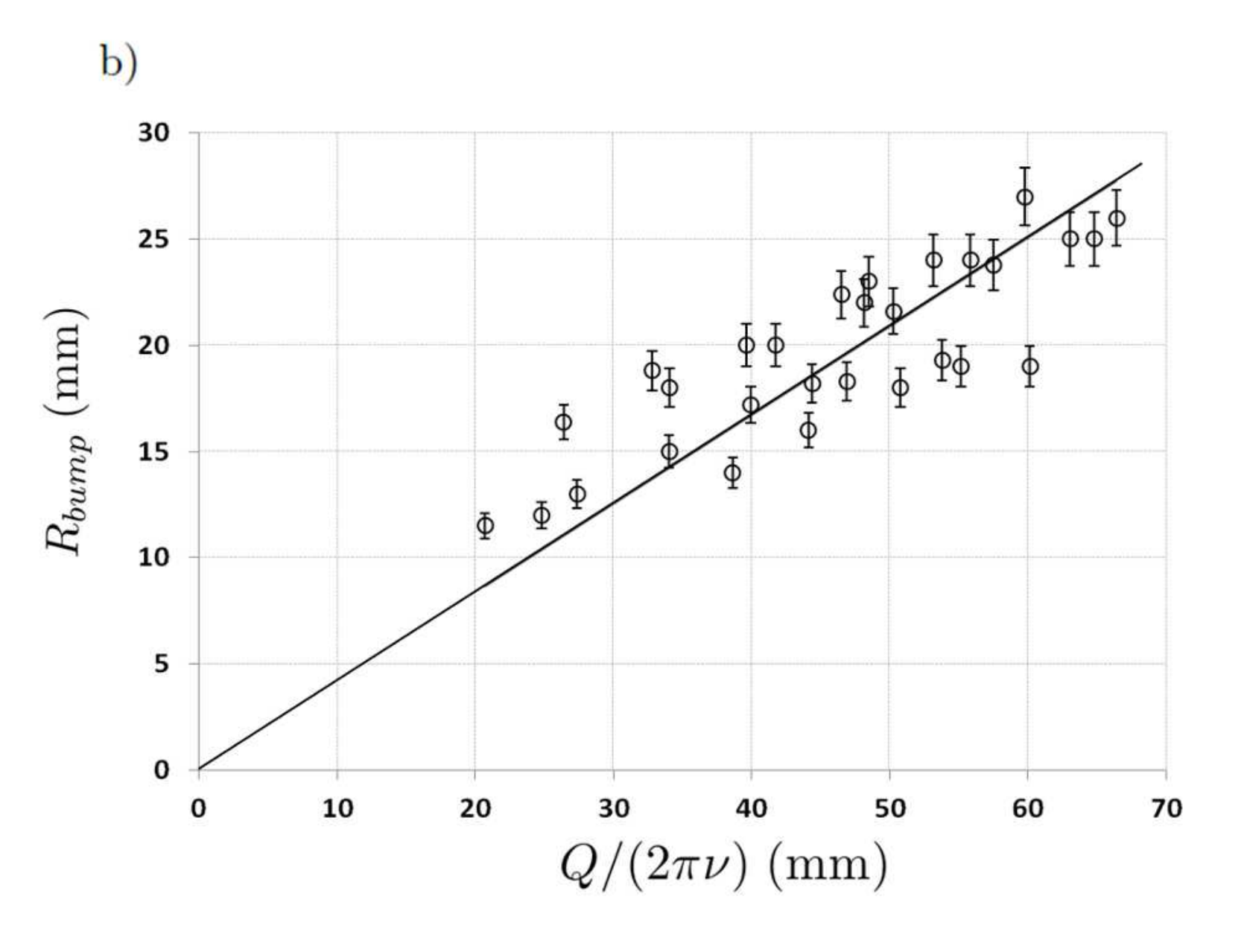}
\caption{
a) The dependence of the circular bump amplitude, $2g\delta H$ (m$^2$s$^{-2}$), on  $v_A^2=Q^2/(4\pi^2 R_n^2 H^2)$ (m$^2$s$^{-2}$). The uncertainties on $\delta H$ are approximately $20\: \%$. b) The observed dependence of the bump radius $R_{bump}$ (mm) on $Q/(2\pi \nu)$ (mm). The bumps are formed with glycerine-water solutions, with viscosity ranging from $58$ to $96$ cS}
\label{scale}
\end{figure*}

The height and radius of the circular bump are readily rationalized via scaling arguments. 
We consider a point $A$ at the surface 
near the plunging jet and a point $B$ on the bump (see Figure \ref{medium}b). 
We denote by $\delta H$ the amplitude of the bump. Since B can be considered as  a stagnation point and since curvature pressures are expected to be negligible with respect to the hydrostatic pressure within the bump, Bernoulli's Theorem dictates that $v_A^2/2 - g\delta H=const.$, so we expect that $\delta H=c_1 v_A^2/2g$ with $v_A\simeq Q/(2\pi R_n H)$.  
Figure \ref{scale}a illustrates the dependence of $2g\delta H$ on $Q^2/(4\pi^2 R_n^2 H^2)$
over the parameter range in which circular bumps arise. This simple scaling is roughly validated,
and a proportionality constant of $c_1=0.41$ is indicated.

Figure \ref{scale}b illustrates the dependence of the bump radius $R_{bump}$ on the flux $Q$ and the kinematic viscosity $\nu$ of the fluid over the range of Weber and Reynolds 
numbers in which circular bumps emerge. The characteristic radius of the inner vortex can be deduced by considering the azimuthal 
component of the vorticity equation. In a steady state, the balance of convection and diffusion of vorticity $\omega$ requires that  
$\left( \mathbf{v}\cdot \bold{\nabla }\right)\mathbf{\omega}\sim\nu \triangle \mathbf{\omega}$. 
The typical scale of the vertical flow is $H_{int}$, the inner depth. 
Thus, balancing $\left( \mathbf{v}\cdot \bold{\nabla }\right)\mathbf{\omega} \sim v\omega/H_{int}$ and $ \nu \triangle \mathbf{\omega}\sim \nu \omega /H_{int}^2$ and using $v\sim Q/\left( 2\pi R_{bump} H_{int}\right) $ indicates that $R_{bump}=c_2 Q/ \left( 2\pi \nu \right)$. Figure~\ref{scale}b lends 
support to this scaling argument, and suggests a proportionality coefficient of $c_2=2.5$.

\begin{figure*}
    \centering
\includegraphics[width=160mm,height=120mm]{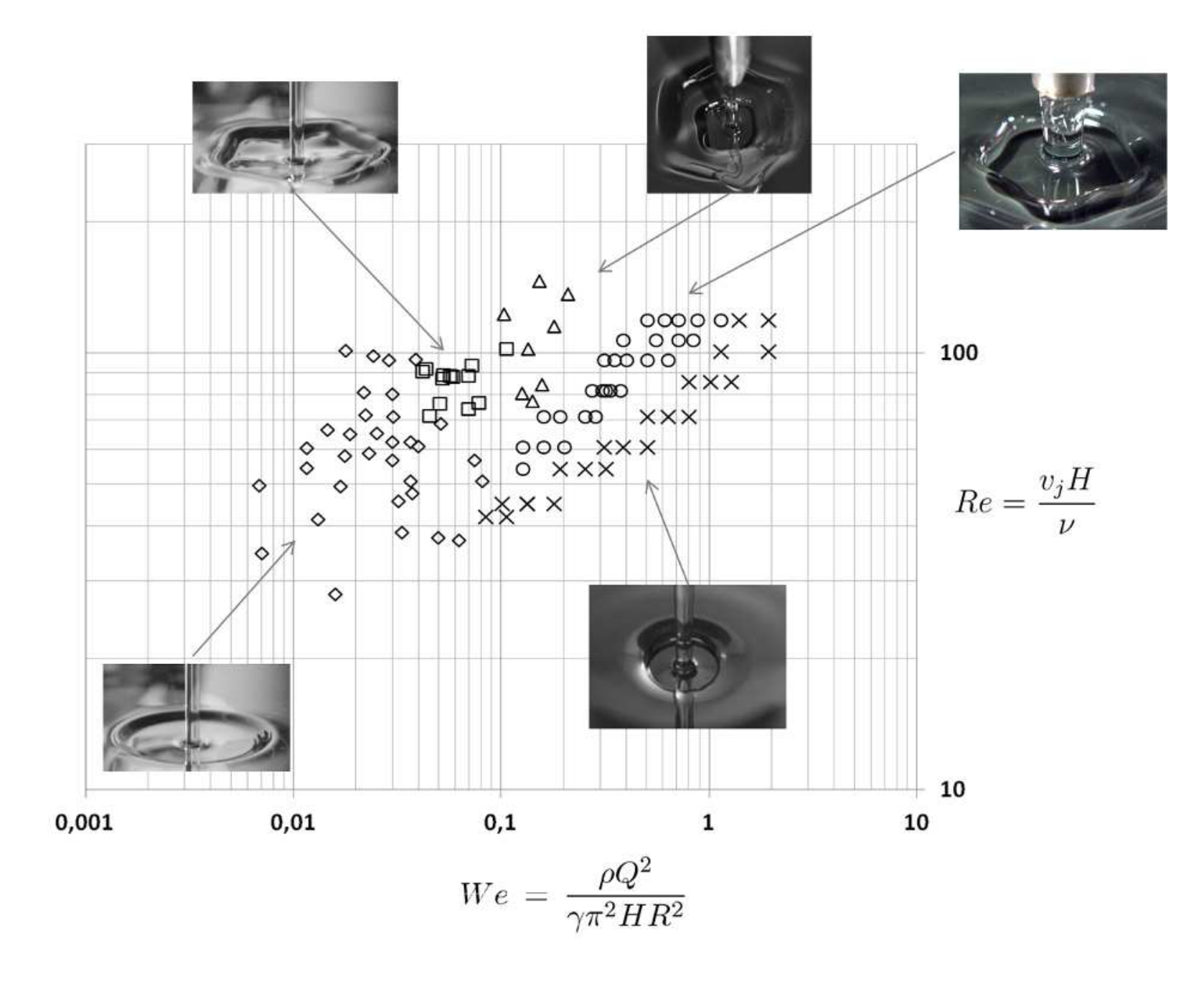}
\caption{The dependence of the flow structure on Reynolds $Re=( v_j H)/\nu$ and Weber number $We=\rho Q^2 /( \gamma \pi^2 H R^2) $, where $v_j=Q/(2\pi R_n)$ is the jet speed, and $R$ the radius of the bump or jump. $\diamond$: circular bumps observed for viscosities in the range of $\nu=58-96$ cS.  $\square$: polygonal bumps ($\nu=58-96$ cS) with number of sides ranging from 5 to 10. $\triangle$: the double jump structure for which a polygonal jump is enclosed by a polygonal bump.  
$\times$: circular type I jumps ($\nu=10$ cS, from Bush \textit{et al.} \cite{Bush2}). $\circ$: polygonal jumps with 3 to 10 sides ($\nu=10$ cS, data from Bush \textit{et al}. \cite{Bush2}). Bush, J.W.M. and Aristoff, J.M. and Hosoi, A.E., J. Fluid Mech. \textbf{558}, (2006) used with permission }
\label{diagrammephase}
\end{figure*}

As illustrated in Figure \ref{bumppicture}c and \ref{bumppicture}i, non-circular bumps may also arise downstream of polygonal hydraulic 
jumps. Similar nested jump-bump structures have been reported by Andersen {\it et al.}\cite{Bohr10} and 
Bush {\it et al.}\cite{Bush2} . We note that the number of sides of the outer bump and inner jump polygons are
not necessarily the same. Figure 1i illustrates a square jump within a pentagonal bump.

Figure~\ref{diagrammephase} indicates where the various flow structures, specifically
circular and polygonal hydraulic jumps and bumps, arise in the 
($We,Re$) plane. In addition to our new data, this regime diagram includes data from Bush \textit{et al.}\cite{Bush2}, with  
due care given to their different definitions of the Weber and Reynolds numbers.
In our experiments, as $Q$ is increased, the data necessarily traverses a path on the regime diagram
along which $Re \propto \sqrt{We}$. The dependence of the flow structure on viscosity was examined by progressively
diluting the solution. We note that in sufficiently dilute solutions, the polygonal forms are unstable owing to the onset of 
turbulence. Viscosity is thus critical in the suppression of turbulence, and the sustenance of stable jumps,
which only arise in the moderate Reynolds number range: $Re \simeq 20-100$.\\

\section{\label{Conclusion}Conclusion}
We have characterized the flows generated by a laminar fluid jet plunging into a bath of the 
same fluid, giving particular attention to the accompanying subsurface vortex structure and its instability.
We have reported and rationalized a new interfacial structure, the circular hydraulic bump, the small surface 
deflection that arises prior to the onset of the hydraulic jump. The bump coincides with the 
stagnation point associated with the subsurface vortex generated by the plunging jet and consists of a small toroidal vortex with a surface signature.
Simple scaling arguments have allowed us to rationalize both the radius and amplitude 
of the bump.\\

We have also reported that, as the flux increases, the circular bump goes unstable to a polygonal
form reminiscent of that arising in the hydraulic jump \cite{Bush2}. We note that the polygonal 
instabilities of both the jump and bump are associated with a toroidal vortex with a surface signature. Since the bump has a relatively small surface signature, we expect its 
accompanying subsurface vorticity to provide the dominant mechanism for its instability. This surface vortex instability mechanism, and its relation to the hydraulic bump, the hydraulic
jump, and the toroidal Leidenfrost vortex~\cite{Perrard}, will be the subject of a theoretical 
investigation to be reported elsewhere.

\begin{acknowledgments}
The authors acknowledge the generous financial support of the National Science Foundation through grant number DMS-0907955. M.L. is grateful to Jos\'{e} Bico, Marc Fermigier from the PMMH laboratory and  
``La soci\'{e}t\'{e} des Amis de l'ESPCI ParisTech'' for funding  this partnership work between the ESPCI ParisTech and MIT. We would like to thank Yves Couder, St\'{e}phane Perrard and Laurent Limat at the MSC laboratory for valuable discussions.
\end{acknowledgments}


\begin{thebibliography}{31}%
\makeatletter
\providecommand \@ifxundefined [1]{%
 \@ifx{#1\undefined}
}%
\providecommand \@ifnum [1]{%
 \ifnum #1\expandafter \@firstoftwo
 \else \expandafter \@secondoftwo
 \fi
}%
\providecommand \@ifx [1]{%
 \ifx #1\expandafter \@firstoftwo
 \else \expandafter \@secondoftwo
 \fi
}%
\providecommand \natexlab [1]{#1}%
\providecommand \enquote  [1]{``#1''}%
\providecommand \bibnamefont  [1]{#1}%
\providecommand \bibfnamefont [1]{#1}%
\providecommand \citenamefont [1]{#1}%
\providecommand \href@noop [0]{\@secondoftwo}%
\providecommand \href [0]{\begingroup \@sanitize@url \@href}%
\providecommand \@href[1]{\@@startlink{#1}\@@href}%
\providecommand \@@href[1]{\endgroup#1\@@endlink}%
\providecommand \@sanitize@url [0]{\catcode `\\12\catcode `\$12\catcode
  `\&12\catcode `\#12\catcode `\^12\catcode `\_12\catcode `\%12\relax}%
\providecommand \@@startlink[1]{}%
\providecommand \@@endlink[0]{}%
\providecommand \url  [0]{\begingroup\@sanitize@url \@url }%
\providecommand \@url [1]{\endgroup\@href {#1}{\urlprefix }}%
\providecommand \urlprefix  [0]{URL }%
\providecommand \Eprint [0]{\href }%
\providecommand \doibase [0]{http://dx.doi.org/}%
\providecommand \selectlanguage [0]{\@gobble}%
\providecommand \bibinfo  [0]{\@secondoftwo}%
\providecommand \bibfield  [0]{\@secondoftwo}%
\providecommand \translation [1]{[#1]}%
\providecommand \BibitemOpen [0]{}%
\providecommand \bibitemStop [0]{}%
\providecommand \bibitemNoStop [0]{.\EOS\space}%
\providecommand \EOS [0]{\spacefactor3000\relax}%
\providecommand \BibitemShut  [1]{\csname bibitem#1\endcsname}%
\let\auto@bib@innerbib\@empty
\bibitem [{\citenamefont {B\'{e}langer}(1841)}]{Belanger}%
  \BibitemOpen
  \bibfield  {author} {\bibinfo {author} {\bibfnamefont {J.}~\bibnamefont
  {B\'{e}langer}},\ }\bibfield  {title} {\enquote {\bibinfo {title} { Notes sur
  l'hydraulique},}\ }\href@noop {} {\bibfield  {journal} {\bibinfo  {journal}
  {\'{E}cole Royale des Ponts et Chauss\'{e}es}\ } (\bibinfo {year}
  {1841})}\BibitemShut {NoStop}%
\bibitem [{\citenamefont {{\noopsort{Rayleigh}}{Lord
  Rayleigh}}(1914)}]{Rayleigh14}%
  \BibitemOpen
  \bibfield  {author} {\bibinfo {author} {\bibnamefont
  {{\noopsort{Rayleigh}}{Lord Rayleigh}}},\ }\bibfield  {title} {\enquote
  {\bibinfo {title} {On the theory of long waves and bores},}\ }\href@noop {}
  {\bibfield  {journal} {\bibinfo  {journal} {Proc. R. Soc. Lond. A}\ }\textbf
  {\bibinfo {volume} {90}},\ \bibinfo {pages} {324--328} (\bibinfo {year}
  {1914})}\BibitemShut {NoStop}%
\bibitem [{\citenamefont {Tani}(1949)}]{Tani48}%
  \BibitemOpen
  \bibfield  {author} {\bibinfo {author} {\bibfnamefont {I.}~\bibnamefont
  {Tani}},\ }\bibfield  {title} {\enquote {\bibinfo {title} {{Water jump in the
  boundary layer}},}\ }\href@noop {} {\bibfield  {journal} {\bibinfo  {journal}
  {{J. Phys. Soc. Jpn.}}\ }\textbf {\bibinfo {volume} {{4}}},\ \bibinfo {pages}
  {{212--215}} (\bibinfo {year} {{1949}})}\BibitemShut {NoStop}%
\bibitem [{\citenamefont {Watson}(1964)}]{Watson}%
  \BibitemOpen
  \bibfield  {author} {\bibinfo {author} {\bibfnamefont {E.}~\bibnamefont
  {Watson}},\ }\bibfield  {title} {\enquote {\bibinfo {title} {{The radial
  spread of a liquid jet over a horizontal plane}},}\ }\href@noop {} {\bibfield
   {journal} {\bibinfo  {journal} {{J. Fluid Mech.}}\ }\textbf {\bibinfo
  {volume} {{20}}},\ \bibinfo {pages} {{481--499}} (\bibinfo {year}
  {{1964}})}\BibitemShut {NoStop}%
\bibitem [{\citenamefont {Bohr}, \citenamefont {Dimon},\ and\ \citenamefont
  {Putkaradze}(1993)}]{Bohr93}%
  \BibitemOpen
  \bibfield  {author} {\bibinfo {author} {\bibfnamefont {T.}~\bibnamefont
  {Bohr}}, \bibinfo {author} {\bibfnamefont {P.}~\bibnamefont {Dimon}}, \ and\
  \bibinfo {author} {\bibfnamefont {V.}~\bibnamefont {Putkaradze}},\ }\bibfield
   {title} {\enquote {\bibinfo {title} {{Shallow-water approach to the circular
  hydraulic jump}},}\ }\href@noop {} {\bibfield  {journal} {\bibinfo  {journal}
  {{J. Fluid Mech.}}\ }\textbf {\bibinfo {volume} {{254}}},\ \bibinfo {pages}
  {{635--648}} (\bibinfo {year} {{1993}})}\BibitemShut {NoStop}%
\bibitem [{\citenamefont {Bohr}\ \emph {et~al.}(1996)\citenamefont {Bohr},
  \citenamefont {Ellegaard}, \citenamefont {Hansen},\ and\ \citenamefont
  {Haaning}}]{Bohr96}%
  \BibitemOpen
  \bibfield  {author} {\bibinfo {author} {\bibfnamefont {T.}~\bibnamefont
  {Bohr}}, \bibinfo {author} {\bibfnamefont {C.}~\bibnamefont {Ellegaard}},
  \bibinfo {author} {\bibfnamefont {A.E.}~\bibnamefont {Hansen}}, \ and\ \bibinfo
  {author} {\bibfnamefont {A.}~\bibnamefont {Haaning}},\ }\bibfield  {title}
  {\enquote {\bibinfo {title} {Hydraulic jumps, flow separation and wave
  breaking: An experimental study},}\ }\href@noop {} {\bibfield  {journal}
  {\bibinfo  {journal} {Physica B}\ }\textbf {\bibinfo {volume} {228}},\
  \bibinfo {pages} {1--10} (\bibinfo {year} {1996})}\BibitemShut {NoStop}%
\bibitem [{\citenamefont {Bohr}, \citenamefont {Putkaradze},\ and\
  \citenamefont {Watanabe}(1997)}]{Bohr97}%
  \BibitemOpen
  \bibfield  {author} {\bibinfo {author} {\bibfnamefont {T.}~\bibnamefont
  {Bohr}}, \bibinfo {author} {\bibfnamefont {V.}~\bibnamefont {Putkaradze}}, \
  and\ \bibinfo {author} {\bibfnamefont {S.}~\bibnamefont {Watanabe}},\
  }\bibfield  {title} {\enquote {\bibinfo {title} {{Averaging theory for the
  structure of hydraulic jumps and separation in laminar free-surface
  flows}},}\ }\href@noop {} {\bibfield  {journal} {\bibinfo  {journal} {{Phys.
  Rev. Lett.}}\ }\textbf {\bibinfo {volume} {{79}}},\ \bibinfo {pages}
  {{1038--1041}} (\bibinfo {year} {{1997}})}\BibitemShut {NoStop}%
\bibitem [{\citenamefont {Bush}\ and\ \citenamefont {Aristoff}(2003)}]{Bush1}%
  \BibitemOpen
  \bibfield  {author} {\bibinfo {author} {\bibfnamefont {J.W.M.}~\bibnamefont
  {Bush}}\ and\ \bibinfo {author} {\bibfnamefont {J.M}~\bibnamefont
  {Aristoff}},\ }\bibfield  {title} {\enquote {\bibinfo {title} {{The influence
  of surface tension on the circular hydraulic jump}},}\ }\href@noop {}
  {\bibfield  {journal} {\bibinfo  {journal} {{J. Fluid Mech.}}\ }\textbf
  {\bibinfo {volume} {{489}}},\ \bibinfo {pages} {{229--238}} (\bibinfo {year}
  {{2003}})}\BibitemShut {NoStop}%
\bibitem [{\citenamefont {Kasimov}(2008)}]{Kasimov08}%
  \BibitemOpen
  \bibfield  {author} {\bibinfo {author} {\bibfnamefont {A.R.}~\bibnamefont
  {Kasimov}},\ }\bibfield  {title} {\enquote {\bibinfo {title} {{A stationary
  circular hydraulic jump, the limits of its existence and its gasdynamic
  analogue}},}\ }\href@noop {} {\bibfield  {journal} {\bibinfo  {journal} {{J.
  Fluid Mech.}}\ }\textbf {\bibinfo {volume} {{601}}},\ \bibinfo {pages}
  {{189--198}} (\bibinfo {year} {{2008}})}\BibitemShut {NoStop}%
\bibitem [{\citenamefont {Watanabe}, \citenamefont {Putkaradze},\ and\
  \citenamefont {Bohr}(2003)}]{Bohr03}%
  \BibitemOpen
  \bibfield  {author} {\bibinfo {author} {\bibfnamefont {S.}~\bibnamefont
  {Watanabe}}, \bibinfo {author} {\bibfnamefont {V.}~\bibnamefont
  {Putkaradze}}, \ and\ \bibinfo {author} {\bibfnamefont {T.}~\bibnamefont
  {Bohr}},\ }\bibfield  {title} {\enquote {\bibinfo {title} {{Integral methods
  for shallow free-surface flows with separation}},}\ }\href@noop {} {\bibfield
   {journal} {\bibinfo  {journal} {{J. Fluid Mech.}}\ }\textbf {\bibinfo
  {volume} {{480}}},\ \bibinfo {pages} {{233--265}} (\bibinfo {year}
  {{2003}})}\BibitemShut {NoStop}%
  \bibitem [{\citenamefont {Andersen}, \citenamefont {Bohr},\ and\ \citenamefont
  {Schnipper}(2009)}]{Bohr10}%
  \BibitemOpen
  \bibfield  {author} {\bibinfo {author} {\bibfnamefont {A.}~\bibnamefont
  {Andersen}}, \bibinfo {author} {\bibfnamefont {T.}~\bibnamefont {Bohr}}, \
  and\ \bibinfo {author} {\bibfnamefont {T.}~\bibnamefont {Schnipper}},\
  }\bibfield  {title} {\enquote {\bibinfo {title} {Separation vortices and
  pattern formation},}\ }\href@noop {} {\bibfield  {journal} {\bibinfo
  {journal} {Theor. Comput. Fluid Dyn.}\ }\textbf {\bibinfo {volume} {24}},\
  \bibinfo {pages} {329--334} (\bibinfo {year} {2009})}\BibitemShut {NoStop}%
  \bibitem [{\citenamefont {Yokoi}\ and\ \citenamefont {Xiao}(2002)}]{Xiao02}%
  \BibitemOpen
  \bibfield  {author} {\bibinfo {author} {\bibfnamefont {K.}~\bibnamefont
  {Yokoi}}\ and\ \bibinfo {author} {\bibfnamefont {F.}~\bibnamefont {Xiao}},\
  }\bibfield  {title} {\enquote {\bibinfo {title} {{Mechanism of structure
  formation in circular hydraulic jumps: numerical studies of strongly deformed
  free-surface shallow flows}},}\ }\href@noop {} {\bibfield  {journal}
  {\bibinfo  {journal} {{Physica D}}\ }\textbf {\bibinfo {volume} {{161}}},\
  \bibinfo {pages} {{202--219}} (\bibinfo {year} {{2002}})}\BibitemShut
  {NoStop}%
\bibitem [{\citenamefont {Liu}\ and\ \citenamefont
  {Lienhard}(1993)}]{Lienhard93}%
  \BibitemOpen
  \bibfield  {author} {\bibinfo {author} {\bibfnamefont {X.}~\bibnamefont
  {Liu}}\ and\ \bibinfo {author} {\bibfnamefont {J.H.}~\bibnamefont {Lienhard~V}},\
  }\bibfield  {title} {\enquote {\bibinfo {title} {{The hydraulic jump in
  circular jet impingement and in other thin liquid films}},}\ }\href@noop {}
  {\bibfield  {journal} {\bibinfo  {journal} {{Exp. Fluids}}\ }\textbf
  {\bibinfo {volume} {{15}}},\ \bibinfo {pages} {{108--116}} (\bibinfo {year}
  {{1993}})}\BibitemShut {NoStop}%
\bibitem [{\citenamefont {Bush}, \citenamefont {Aristoff},\ and\ \citenamefont
  {Hosoi}(2006)}]{Bush2}%
  \BibitemOpen
  \bibfield  {author} {\bibinfo {author} {\bibfnamefont {J.W.M.}~\bibnamefont
  {Bush}}, \bibinfo {author} {\bibfnamefont {J.M.}~\bibnamefont {Aristoff}}, \
  and\ \bibinfo {author} {\bibfnamefont {A.E.}~\bibnamefont {Hosoi}},\ }\bibfield
   {title} {\enquote {\bibinfo {title} {{An experimental investigation of the
  stability of the circular hydraulic jump}},}\ }\href@noop {} {\bibfield
  {journal} {\bibinfo  {journal} {{J. Fluid Mech.}}\ }\textbf {\bibinfo
  {volume} {{558}}},\ \bibinfo {pages} {{33--52}} (\bibinfo {year}
  {{2006}})}\BibitemShut {NoStop}%
\bibitem [{\citenamefont {Ellegaard}\ \emph {et~al.}(1998)\citenamefont
  {Ellegaard}, \citenamefont {Hansen}, \citenamefont {Haaning}, \citenamefont
  {Hansen}, \citenamefont {Marcussen}, \citenamefont {Bohr}, \citenamefont
  {Hansen},\ and\ \citenamefont {Watanabe}}]{Bohr98}%
  \BibitemOpen
  \bibfield  {author} {\bibinfo {author} {\bibfnamefont {C.}~\bibnamefont
  {Ellegaard}}, \bibinfo {author} {\bibfnamefont {A.}~\bibnamefont {Hansen}},
  \bibinfo {author} {\bibfnamefont {A.}~\bibnamefont {Haaning}}, \bibinfo
  {author} {\bibfnamefont {K.}~\bibnamefont {Hansen}}, \bibinfo {author}
  {\bibfnamefont {A.}~\bibnamefont {Marcussen}}, \bibinfo {author}
  {\bibfnamefont {T.}~\bibnamefont {Bohr}}, \bibinfo {author} {\bibfnamefont
  {J.}~\bibnamefont {Hansen}}, \ and\ \bibinfo {author} {\bibfnamefont
  {S.}~\bibnamefont {Watanabe}},\ }\bibfield  {title} {\enquote {\bibinfo
  {title} {{Creating corners in kitchen sinks}},}\ }\href@noop {} {\bibfield
  {journal} {\bibinfo  {journal} {{Nature}}\ }\textbf {\bibinfo {volume}
  {{392}}},\ \bibinfo {pages} {{767--768}} (\bibinfo {year}
  {{1998}})}\BibitemShut {NoStop}%
\bibitem [{\citenamefont {Ellegaard}\ \emph {et~al.}(1999)\citenamefont
  {Ellegaard}, \citenamefont {Hansen}, \citenamefont {Haaning}, \citenamefont
  {Hansen}, \citenamefont {Marcussen}, \citenamefont {Bohr}, \citenamefont
  {Hansen},\ and\ \citenamefont {Watanabe}}]{Bohr99}%
  \BibitemOpen
  \bibfield  {author} {\bibinfo {author} {\bibfnamefont {C.}~\bibnamefont
  {Ellegaard}}, \bibinfo {author} {\bibfnamefont {A.E.}~\bibnamefont {Hansen}},
  \bibinfo {author} {\bibfnamefont {A.}~\bibnamefont {Haaning}}, \bibinfo
  {author} {\bibfnamefont {K.}~\bibnamefont {Hansen}}, \bibinfo {author}
  {\bibfnamefont {A.}~\bibnamefont {Marcussen}}, \bibinfo {author}
  {\bibfnamefont {T.}~\bibnamefont {Bohr}}, \bibinfo {author} {\bibfnamefont
  {J.L.}~\bibnamefont {Hansen}}, \ and\ \bibinfo {author} {\bibfnamefont
  {S.}~\bibnamefont {Watanabe}},\ }\bibfield  {title} {\enquote {\bibinfo
  {title} {{Cover illustration: polygonal hydraulic jumps}},}\ }\href@noop {}
  {\bibfield  {journal} {\bibinfo  {journal} {{Nonlinearity}}\ }\textbf
  {\bibinfo {volume} {{12}}},\ \bibinfo {pages} {{1--7}} (\bibinfo {year}
  {{1999}})}\BibitemShut {NoStop}%
\bibitem [{\citenamefont {Martens}, \citenamefont {Watanabe},\ and\
  \citenamefont {Bohr}(2012)}]{Bohr12}%
  \BibitemOpen
  \bibfield  {author} {\bibinfo {author} {\bibfnamefont {E.}~\bibnamefont
  {Martens}}, \bibinfo {author} {\bibfnamefont {S.}~\bibnamefont {Watanabe}}, \
  and\ \bibinfo {author} {\bibfnamefont {T.}~\bibnamefont {Bohr}},\ }\bibfield
  {title} {\enquote {\bibinfo {title} {Model for polygonal hydraulic jumps},}\
  }\href@noop {} {\bibfield  {journal} {\bibinfo  {journal} {Phys. Rev. E}\
  }\textbf {\bibinfo {volume} {85}} (\bibinfo {year} {2012})}\BibitemShut
  {NoStop}%
\bibitem [{\citenamefont {Plateau}(1873)}]{Plateau73}%
  \BibitemOpen
  \bibfield  {author} {\bibinfo {author} {\bibfnamefont {J.}~\bibnamefont
  {Plateau}},\ }\bibfield  {title} {\enquote {\bibinfo {title} {Experimental
  and theoretical statics of liquids subject to molecular forces only},}\
  }\href@noop {} {\bibfield  {journal} {\bibinfo  {journal} {Gauthier-Villars,
  Paris}\ }\textbf {\bibinfo {volume} {1}} ,\
  \bibinfo {pages} {4--13}(\bibinfo {year}
  {1873})}\BibitemShut {NoStop}%
\bibitem [{\citenamefont {{\noopsort{Rayleigh}}{Lord
  Rayleigh}}(1878)}]{Rayleigh79}%
  \BibitemOpen
  \bibfield  {author} {\bibinfo {author} {\bibnamefont
  {{\noopsort{Rayleigh}}{Lord Rayleigh}}},\ }\bibfield  {title} {\enquote
  {\bibinfo {title} {On the instability of jets},}\ }\href@noop {} {\bibfield
  {journal} {\bibinfo  {journal} {Proc. Lond. M. Soc.}\ }\textbf {\bibinfo
  {volume} {10}} (\bibinfo {year} {1878})}\BibitemShut {NoStop}%
\bibitem [{\citenamefont {Hocking}\ and\ \citenamefont
  {Michael}(1959)}]{Hocking59}%
  \BibitemOpen
  \bibfield  {author} {\bibinfo {author} {\bibfnamefont {L.~M.}\ \bibnamefont
  {Hocking}}\ and\ \bibinfo {author} {\bibfnamefont {D.~H.}\ \bibnamefont
  {Michael}},\ }\bibfield  {title} {\enquote {\bibinfo {title} {The stability
  of a column of rotating liquid},}\ }\href@noop {} {\bibfield  {journal}
  {\bibinfo  {journal} {Mathematika}\ }\textbf {\bibinfo {volume} {6}},\
  \bibinfo {pages} {25--32} (\bibinfo {year} {1959})}\BibitemShut {NoStop}%
\bibitem [{\citenamefont {Pedley}(1967)}]{Pedley1}%
  \BibitemOpen
  \bibfield  {author} {\bibinfo {author} {\bibfnamefont {T.J.}~\bibnamefont
  {Pedley}},\ }\bibfield  {title} {\enquote {\bibinfo {title} {The stability of
  rotating flows with a cylindrical free surface},}\ }\href@noop {} {\bibfield
  {journal} {\bibinfo  {journal} {J. Fluid Mech.}\ }\textbf {\bibinfo {volume}
  {30}},\ \bibinfo {pages} {127--147} (\bibinfo {year} {1967})}\BibitemShut
  {NoStop}%
\bibitem [{\citenamefont {Kubitschek}\ and\ \citenamefont
  {Weidman}(2007)}]{Weidman06}%
  \BibitemOpen
  \bibfield  {author} {\bibinfo {author} {\bibfnamefont {J.}~\bibnamefont
  {Kubitschek}}\ and\ \bibinfo {author} {\bibfnamefont {P.}~\bibnamefont
  {Weidman}},\ }\bibfield  {title} {\enquote {\bibinfo {title} {The effect of
  viscosity on the stability of a uniformly rotating liquid column in zero
  gravity},}\ }\href@noop {} {\bibfield  {journal} {\bibinfo  {journal} {J.
  Fluid Mech.}\ }\textbf {\bibinfo {volume} {572}},\ \bibinfo {pages} {261--286} (\bibinfo {year}
  {2007})}\BibitemShut {NoStop}%
\bibitem [{\citenamefont {{\noopsort{Kelvin}}{Lord Kelvin}}(1867)}]{Kelvin67}%
  \BibitemOpen
  \bibfield  {author} {\bibinfo {author} {\bibfnamefont {Sir W.}~\bibnamefont
  {Thomson}},\ }\bibfield  {title} {\enquote {\bibinfo
  {title} {{On vortex atoms}},}\ }\href@noop {} {\bibfield  {journal} {\bibinfo
   {journal} {{Phil. Mag.}}\ }\textbf {\bibinfo {volume} {{34}}},\ \bibinfo
  {pages} {{15 -- 24}} (\bibinfo {year} {{1867}})}\BibitemShut {NoStop}%
\bibitem [{\citenamefont {Thomson}(1883)}]{Thomson83}%
  \BibitemOpen
  \bibfield  {author} {\bibinfo {author} {\bibfnamefont {J.J.}~\bibnamefont
  {Thomson}},\ }\bibfield  {title} {\enquote {\bibinfo {title} {{A treatise on
  the motion of vortex ring}},}\ }\href@noop {} {\bibfield  {journal} {\bibinfo
   {journal} {Macmillan, London}\ } (\bibinfo {year} {{1883}})}\BibitemShut
  {NoStop}%
\bibitem [{\citenamefont {Widnall}\ and\ \citenamefont
  {Tsai}(1977)}]{Widnall77}%
  \BibitemOpen
  \bibfield  {author} {\bibinfo {author} {\bibfnamefont {S.E.}~\bibnamefont
  {Widnall}}\ and\ \bibinfo {author} {\bibfnamefont {C.Y.}~\bibnamefont {Tsai}},\
  }\bibfield  {title} {\enquote {\bibinfo {title} {The instability of the thin
  vortex ring of constant vorticity},}\ }\href@noop {} {\bibfield  {journal}
  {\bibinfo  {journal} {Phil. Trans. R. Soc. Lond. A}\ }\textbf {\bibinfo
  {volume} {{287}}},\ \bibinfo {pages} {{273--305}} (\bibinfo {year}
  {{1977}})}\BibitemShut {NoStop}%
\bibitem [{\citenamefont {Maxworthy}(1977)}]{Maxworthy77}%
  \BibitemOpen
  \bibfield  {author} {\bibinfo {author} {\bibfnamefont {T.}~\bibnamefont
  {Maxworthy}},\ }\bibfield  {title} {\enquote {\bibinfo {title} {{Some
  experimental studies of vortex rings}},}\ }\href@noop {} {\bibfield
  {journal} {\bibinfo  {journal} {{J. Fluid Mech.}}\ }\textbf {\bibinfo
  {volume} {{81}}},\ \bibinfo {pages} {{465 -- 495}} (\bibinfo {year}
  {{1977}})}\BibitemShut {NoStop}%
\bibitem [{\citenamefont {Saffman}(1978)}]{Saffman78}%
  \BibitemOpen
  \bibfield  {author} {\bibinfo {author} {\bibfnamefont {P.}~\bibnamefont
  {Saffman}},\ }\bibfield  {title} {\enquote {\bibinfo {title} {The number of
  waves on unstable vortex rings},}\ }\href@noop {} {\bibfield  {journal}
  {\bibinfo  {journal} {J. Fluid Mech.}\ }\textbf {\bibinfo {volume} {84}},\
  \bibinfo {pages} {625 -- 639} (\bibinfo {year} {1978})}\BibitemShut {NoStop}%
\bibitem [{\citenamefont {Knio}\ and\ \citenamefont
  {Ghoniem}(1990)}]{Ghoniem90}%
  \BibitemOpen
  \bibfield  {author} {\bibinfo {author} {\bibfnamefont {O.M.}~\bibnamefont
  {Knio}}\ and\ \bibinfo {author} {\bibfnamefont {A.F.}~\bibnamefont {Ghoniem}},\
  }\bibfield  {title} {\enquote {\bibinfo {title} {Numerical study of a
  three-dimensional vortex method},}\ }\href@noop {} {\bibfield  {journal}
  {\bibinfo  {journal} {J. Comput. Phys.}\ }\textbf {\bibinfo {volume} {86}},\
  \bibinfo {pages} {75--106} (\bibinfo {year} {1990})}\BibitemShut {NoStop}%
\bibitem [{\citenamefont {Perrard}\ \emph {et~al.}(2012)\citenamefont
  {Perrard}, \citenamefont {Couder}, \citenamefont {Fort},\ and\ \citenamefont
  {Limat}}]{Perrard}%
  \BibitemOpen
  \bibfield  {author} {\bibinfo {author} {\bibfnamefont {S.}~\bibnamefont
  {Perrard}}, \bibinfo {author} {\bibfnamefont {Y.}~\bibnamefont {Couder}},
  \bibinfo {author} {\bibfnamefont {E.}~\bibnamefont {Fort}}, \ and\ \bibinfo
  {author} {\bibfnamefont {L.}~\bibnamefont {Limat}},\ }\bibfield  {title}
  {\enquote {\bibinfo {title} {Leidenfrost levitated liquid tori},}\
  }\href@noop {} {\bibfield  {journal} {\bibinfo  {journal} {Europhys. Lett.}\
  }\textbf {\bibinfo {volume} {100}} ,\
  \bibinfo {pages} {54006} (\bibinfo {year} {{2012}})}\BibitemShut
  {NoStop}%
\bibitem [{\citenamefont {Bin}(1993)}]{Bin93}%
  \BibitemOpen
  \bibfield  {author} {\bibinfo {author} {\bibfnamefont {A.K}~\bibnamefont
  {Bi\'{n}}},\ }\bibfield  {title} {\enquote {\bibinfo {title} {Gas entrainment by
  plunging liquid jets},}\ }\href@noop {} {\bibfield  {journal} {\bibinfo
  {journal} {Chem. Eng. Sci.}\ }\textbf {\bibinfo {volume} {48}},\ \bibinfo
  {pages} {3585 -- 3630} (\bibinfo {year} {1993})}\BibitemShut {NoStop}%
\bibitem [{\citenamefont {Lorenceau}, \citenamefont {Qu\'{e}r\'{e}},\ and\
  \citenamefont {Eggers}(2004)}]{Quere04}%
  \BibitemOpen
  \bibfield  {author} {\bibinfo {author} {\bibfnamefont {E.}~\bibnamefont
  {Lorenceau}}, \bibinfo {author} {\bibfnamefont {D.}~\bibnamefont
  {Qu\'{e}r\'{e}}}, \ and\ \bibinfo {author} {\bibfnamefont {J.}~\bibnamefont
  {Eggers}},\ }\bibfield  {title} {\enquote {\bibinfo {title} {{Air entrainment
  by a viscous jet plunging into a bath}},}\ }\href@noop {} {\bibfield
  {journal} {\bibinfo  {journal} {{Phys. Rev. Lett.}}\ }\textbf {\bibinfo
  {volume} {{93}}} (\bibinfo {year} {{2004}})}\BibitemShut {NoStop}%
\end{thebibliography}

\providecommand{\noopsort}[1]{}\providecommand{\singleletter}[1]{#1}%

\end{document}